\documentclass[%
 preprint,
superscriptaddress,
 amsmath,amssymb,
 aip,apl
]{revtex4-1}
\usepackage{url}
\usepackage{color}
\usepackage{graphicx}
\usepackage{epstopdf}
\usepackage{commath}

\begin{document}
\author{T. Elo}
\affiliation{Low Temperature Laboratory, Department of Applied Physics, Aalto University School of Science, P.O. Box 15100, FI-00076 AALTO, Finland}
\author{T. S. Abhilash}
\affiliation{Low Temperature Laboratory, Department of Applied Physics, Aalto University School of Science, P.O. Box 15100, FI-00076 AALTO, Finland}
\author{M. R. Perelshtein}
\affiliation{Low Temperature Laboratory, Department of Applied Physics, Aalto University School of Science, P.O. Box 15100, FI-00076 AALTO, Finland}
\affiliation{Moscow Institute of Physics and Technology, 141700, Russian Federation}
\author{I. Lilja}
\affiliation{Low Temperature Laboratory, Department of Applied Physics, Aalto University School of Science, P.O. Box 15100, FI-00076 AALTO, Finland}
\author{E. V. Korostylev}
\affiliation{Moscow Institute of Physics and Technology, 141700, Russian Federation}
\author{P. J. Hakonen}
\affiliation{Low Temperature Laboratory, Department of Applied Physics, Aalto University School of Science, P.O. Box 15100, FI-00076 AALTO, Finland}

\title{Broadband lumped-element Josephson parametric amplifier with single-step lithography}

\date{\today}

\begin{abstract}
We present a lumped-element Josephson parametric amplifier (JPA) utilizing a straightforward fabrication process involving a single electron beam lithography step followed by double-angle evaporation of aluminum and in-situ oxidation. The Josephson junctions forming the SQUID are fabricated using bridgeless shadow evaporation technique, which enables reliable fabrication of relatively large ($\sim9~\mathrm{\mu m^2}$) junctions. Our strongly coupled flux-pumped JPA achieves 20~dB gain with 95~MHz bandwidth around 5~GHz, while the center frequency is tunable by more than 1~GHz, with the additional possibility for rapid tuning by varying the pump frequency alone. Analytical calculations based on the input-output theory reproduce our measurement results closely.
\end{abstract}

\maketitle

%\section{Introduction}
Low-noise amplification of microwave signals is a key requirement in numerous experiments in quantum technology, including qubit readout, optomechanics and quantum sensors. Since current state-of-the-art semiconductor amplifiers add 1\,--\,3~K %(4\,--\,12~photons at 5~GHz)
of noise to the measured signal \cite{Schleeh2012}, Josephson parametric amplifiers (JPAs) \cite{Zimmer1967,Yurke1988}, along with other superconducting amplifiers \cite{HoEom2012,Lahteenmaki2012,Lahteenmaki2014,Jebari2018}, having added noise close to one quantum have gained significant interest. The active component in a JPA is the nonlinear inductance of a Josephson junction (JJ), modulating the resonance frequency of a resonator, in first realizations formed by a transmission line cavity \cite{Yamamoto2008, Castellanos-Beltran2008, Bergeal2010,Simoen2015a}.
However, the bandwidth of cavity-based JPAs is limited to a few MHz due to the high quality ($Q$) factor of the cavity.

To achieve higher bandwidths, a lumped-element JPA was introduced \cite{Hatridge2011}, where the JJ and a capacitor form a parallel LC resonator, maximizing the ratio of Josephson inductance to the total inductance. The behavior of a lumped element JPA can be tuned over a wide range by changing the coupling strength. Besides high coupled $Q$ configuration\cite{Zhou2014}, one can omit the coupling capacitor resulting in a low coupled $Q$ and wide bandwidth \cite{Hatridge2011, Mutus2013} of up to 100~MHz,  which can be increased further by engineering the impedance of the environment \cite{Mutus2014b, Roy2015a}.
However, these JPAs are relatively complicated to fabricate, requiring deposition of low-loss dielectric for parallel plate capacitor and low-impedance vias \cite{Mutus2013} or optical lithography for niobium followed by e-beam lithography for two aluminum layers \cite{Zhou2014}.

We present a lumped-element JPA utilizing a straightforward fabrication process. The JJs, capacitor, flux line and bonding pads are defined in a single e-beam lithography followed by double-angle evaporation of aluminum. We employ a bridgeless shadow evaporation technique \cite{Lecocq2011} allowing us to fabricate JJs with high critical currents by increasing the surface area instead of lowering oxide thickness, which could lead to increased variance of critical currents.

%{\color{red}\section{Theory}\label{theory}}
 First, we consider the theoretical framework of our lumped-element JPA, operating at frequency $\omega$, consisting of a non-linear resonator formed by a shunt capacitor and a SQUID which is pumped with external RF magnetic flux through the SQUID at frequency $\omega_p \approx 2\omega$ (three-wave mixing). We chose this operation regime, because for a four-wave mixing (typically with a current pump with $\omega_p \approx \omega$) the large amplitude pump is within the amplification bandwidth, whereas the three-wave mixing process separates the pump tone from the amplified signals, therefore simplifying the practical use of the JPA. We can write down the Hamiltonian of a system under study in the rotating wave approximation \cite{Lahteenmaki2013}:
	
	\begin{equation}
		\hat{H}=\hbar \xi_{r}\hat{a}^{\dagger}\hat{a}-\frac{\hbar}{2} \left[\alpha^*\hat{a}^2+\alpha\hat{a}^{\dagger 2} \right]. \\[5pt] \label{H_JPA}
	\end{equation}
	
	\noindent In this formula $\xi_r=\omega_p/2-\omega_r$, where $\omega_r$ is the resonator frequency, $\alpha$ is the strength of pump and $\hat{a}$ is the cavity mode. Solving the Quantum Langevin Equation (QLE) with the Hamiltonian above yields the signal gain as a function of frequency. In QLE we take into consideration a cavity decay rate $\kappa$, which is determined by the JPA quality factor $Q$: $\kappa=\omega/Q$. Since our JPA is strongly coupled to the environment, its quality factor is dominated by coupling, and therefore $Q \approx Q_c$ ($= Z_0 \sqrt{C/L}$\,). The resulting amplification of the signal at frequency $\omega$ as a function of detuning $\xi = \omega_p/2-\omega$ is given by
	
	\begin{equation}
		G(\xi)=\abs{1-\frac{\kappa \chi(\frac{\omega_p}{2}+\xi)}{1-\abs{\alpha}^2\chi\left(\frac{\omega_p}{2}+\xi\right) \chi^{*}\left( \frac{\omega_p}{2}-\xi \right)}}^2, \\[5pt] \label{gain}
	\end{equation}
		
	\noindent where $\chi(\omega)=\left[\kappa/2-i(\omega-\omega_{r})\right]^{-1}$ is the electrical susceptibility of resonator. It can be seen from the equation that gain is achieved at various resonator detunings, but to clarify this dependence we fix the pump power to $\alpha_{\mathrm{max}} = \kappa/2$, corresponding to infinity gain at zero detuning. Now, by substituting $\chi = (\kappa/2 - i\sigma \kappa)^{-1}$, where we introduce $\sigma = \xi_r/\kappa$, the maximum gain at a given resonator detuning is given by:

	\begin{equation}\label{sigma}
	G\,(\sigma)=\frac{(1 + 2\sigma^2)^2}{ 4 \sigma^2\,(1 + \sigma^2)}.
	\end{equation}
	
    \noindent In the high-gain limit we can assume that $\xi_r \ll \kappa$, and in terms of quality factor this relation can be written as $G = |2\,Q\,\xi_r/\omega_r|^{-2}$. Consequently, we can express the pump frequency range where the gain exceeds a given value $G_0$:

    \begin{equation}\label{eq_gain_bw}
    \omega_p (G_0) = \,\omega_r\left(2 \pm \frac{1}{\,\sqrt{G_0}\,Q}\right).
    \end{equation}

    \noindent Then, in addition to wide bandwidth, JPA with low quality factor has the advantage of wide tunability by varying the pump frequency. This tuning via $\omega_p$ is a considerably faster operation than changing the DC flux value. In general, when the pump power is not fixed it is possible to use $\eta = \alpha/\alpha_{\mathrm{max}}=2\,\alpha/\kappa$ to evaluate the final maximum gain function of the two parameters controlled in the experiment:

    \begin{equation}
   G(\sigma, \eta)\,=\frac{\left(\eta ^2+4 \sigma ^2+1\right)^2}{\eta ^4+\eta ^2 \left(8 \sigma ^2-2\right)+\left(4 \sigma ^2+1\right)^2},
    \end{equation}

\noindent where $\eta \leq 1$. Here if $\eta = 1$ this equation turns into Eq.~(\ref{sigma}). Note that in the case $\sigma = 0$, meaning zero resonator detuning, gain tends to infinity.

\begin{figure}[!ht]
\centering
\includegraphics[width=0.95\linewidth]{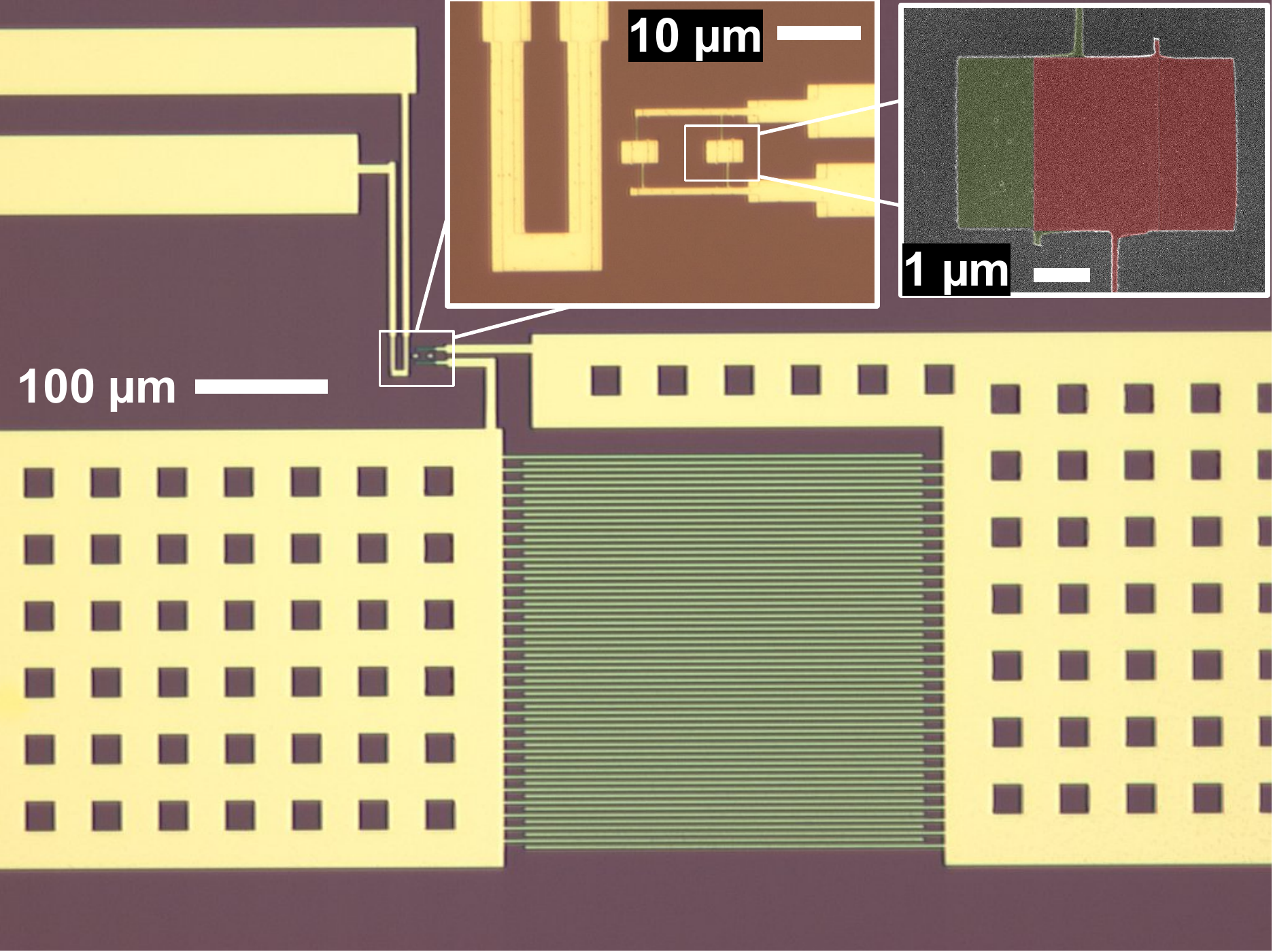}
\caption{Micrographs of a fabricated JPA. Main figure and top-middle inset: Optical micrographs of the JPA. Top-right inset: false-color SEM image of a fabricated Josephson junction with green and red colors denoting first and second aluminum layers, respectively \label{fig-mask}}
\end{figure}

%\section{Design and fabrication}\label{fab}
%Let  us  now  discuss our amplifier design and fabrication process.
The design of our JPA is shown in Fig.~\ref{fig-mask}. The layout design follows a lumped-element principle: no transmission lines are required since all distances are kept short compared to the wavelength ($\lambda \approx 30~\mathrm{mm}$ at 10~GHz). Instead of a ground plane spreading over the entire chip, we place the SQUID and the capacitor in proximity to two bonding pads. The JPA is tuned and pumped with an external AC+DC flux through the SQUID loop, applied with a loop whose geometry is designed to reduce parasitic coupling to the loop formed by the SQUID and the capacitor.

While the resonance frequency of the JPA is given by $\omega_p = 1/\sqrt{L C}$, the amplification bandwidth is inversely proportional to quality factor $Q = Z_0 \sqrt{C/L}$ \cite{Mutus2014b, Roy2015a}, and thus related to critical current of the SQUID as $1/I_c$.
On the other hand, the saturation power of the JPA is directly proportional to $I_c$, resulting in a trade-off between bandwidth and saturation power. Moreover, JPAs with low $Q$ require strong pumping which may cause unwanted nonlinear behavior \cite{Roy2015a,Manucharyan2007}. However, in our application wide bandwidth is preferred over high saturation power, and therefore we set $Q \approx 2$, resulting in $C$ = 1.2~pF and $L$ = 0.8~nH, obtained by applying DC magnetic flux of $\Phi_{\mathrm{DC}} \approx 0.4\,\Phi_0$ through a SQUID with total unbiased $I_c = 1.2~\mu$A.

The capacitor was realized as an interdigital capacitor due to its low losses \cite{Bahl2003} and good suitability to our e-beam lithography process. The capacitor is oriented so that the rotational axis of evaporation angles is perpendicular to the capacitor fingers, allowing proper control over the finger widths. The 1.2~pF capacitor has 62 fingers with 330~$\mathrm{\mu}$m length and 2.4~$\mathrm{\mu}$m width and gap, connected to the bonding pads directly, thus minimizing parasitic inductance. Since large interdigital capacitors may exhibit nonideal behavior at high frequencies, we verified our capacitor design with electromagnetic simulations \footnote{CST Microwave Studio and Sonnet were used for EM simulations.}. %{\color{red} Due to an enhanced per unit area capacitance (12.2 pF/mm$^2$), geometric inductance can be considered adequate small.}
In addition, EM simulations were used to estimate the total geometric inductance of the JPA device, resulting in less than 10~\% of the total inductance at the operating range.

For the Josephson junctions we chose a small critical current density (7~A/$\mathrm{cm^2}$) to improve the quality and reproducibility of the JJs.  This, combined with the 600~nA critical current of the JJs results in junction area of $\sim9~\mathrm{\mu m^2}$. Since such JJs are difficult to fabricate with Dolan bridge technique, we employed a bridgeless shadow evaporation process \cite{Lecocq2011}, where the layer separation is realized in the leads connecting the JJs: the lines leading up- and downwards from the JJ in Fig.~\ref{fig-mask} are formed by the first and second aluminum layers, respectively.

The JPAs were fabricated on oxidized silicon substrate with e-beam lithography using double layer resist  (Copolymer MMA-MAA 700~nm + PMMA 200~nm) and 100~kV beam to achieve low parasitic undercuts. The process involved a double-angle ($\pm 45^\circ$) evaporation of 75~nm thick aluminum, separated by in-situ oxidation at 15~mbar for 10~min.

%\section{Experimental results}

\begin{figure}[!ht]
\centering
\includegraphics[width=0.95\linewidth]{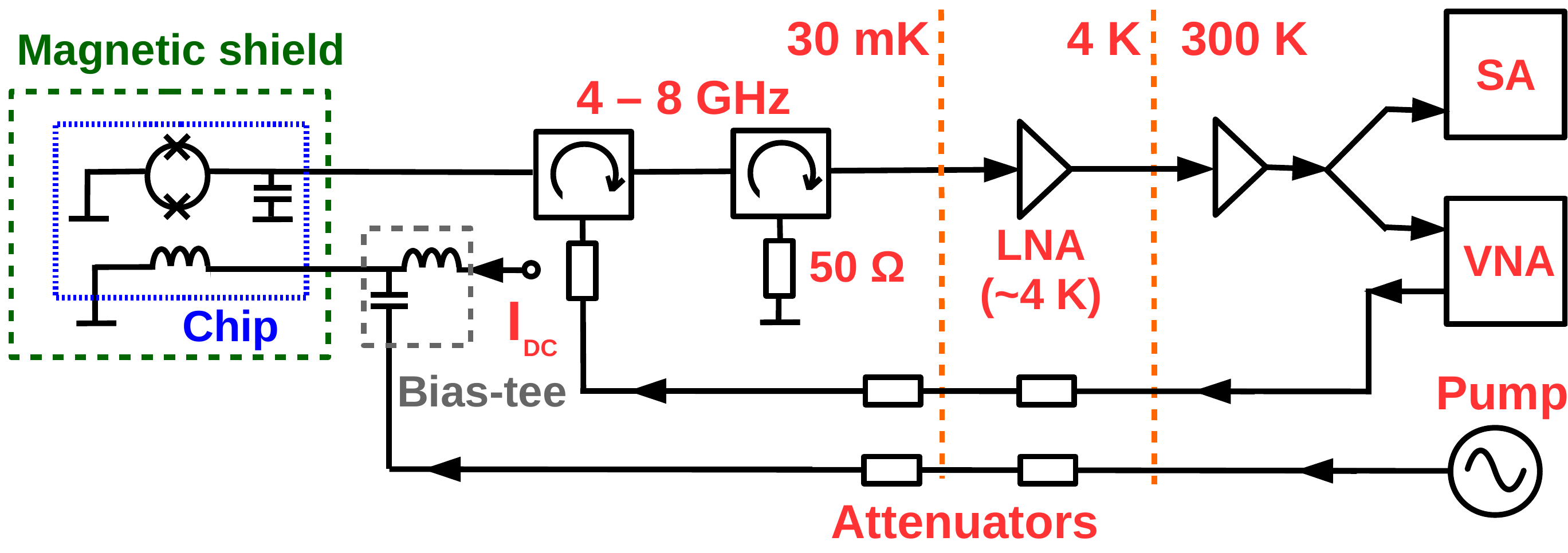}
\caption{Simplified schematic of the measurement system used in characterization of the JPA. The transmission measurements were conducted using a vector network analyzer (VNA) while spectrum analyzer (SA) was used for noise measurements.  \label{fig-meas_schem}}
\end{figure}

Our measurement setup is illustrated in Fig.~\ref{fig-meas_schem}. The measurements were conducted at 30~mK in BlueFors LD-250 dry dilution cryostat. The JPA was protected from external magnetic fields with a cylindrical Cryo\-perm shield and a superconducting (Pb) inner shield. The DC flux bias and RF pump shared a common on-chip flux line, and the signals were combined by an external bias-tee.

\begin{figure}[!ht]
\centering
\includegraphics[width=0.9\linewidth]{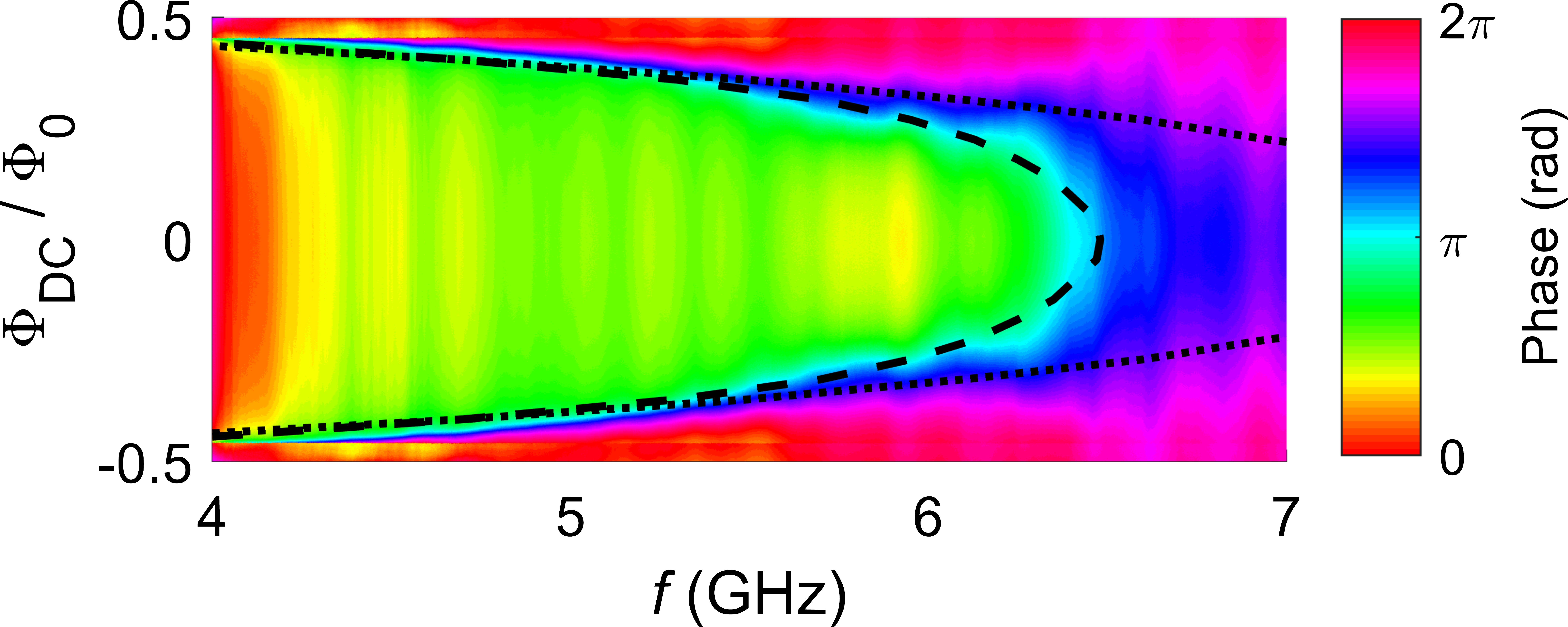}
\caption{Phase of reflected signal as a function of frequency and applied DC flux. The dotted and dashed lines are estimated JPA resonances using ideal and EM-simulated dispersive capacitor models, respectively.  \label{fig-flux}}
\end{figure}

%\section{Results}\label{res}
First, we characterized the tunability of the JPA resonance frequency as a function of DC magnetic flux applied through the SQUID loop.
%Low internal losses and low quality factor of our JPA cause the resonance to disappear at certain frequencies in amplitude plot, and therefore Fig.~\ref{fig-flux} shows the phase of the reflected signal instead.
The Fig.~\ref{fig-flux} shows the phase of the reflected signal and the calculated estimates for resonance frequency using both ideal and EM-simulated capacitor model including parasitic inductance. Although our capacitor contains parasitics, we observe that at the desired working point at 5~GHz the deviation between the two capacitor models is negligible, meaning that the capacitor can be considered nearly ideal at those frequencies.

\begin{figure}[!ht]
\centering
\includegraphics[width=0.9\linewidth]{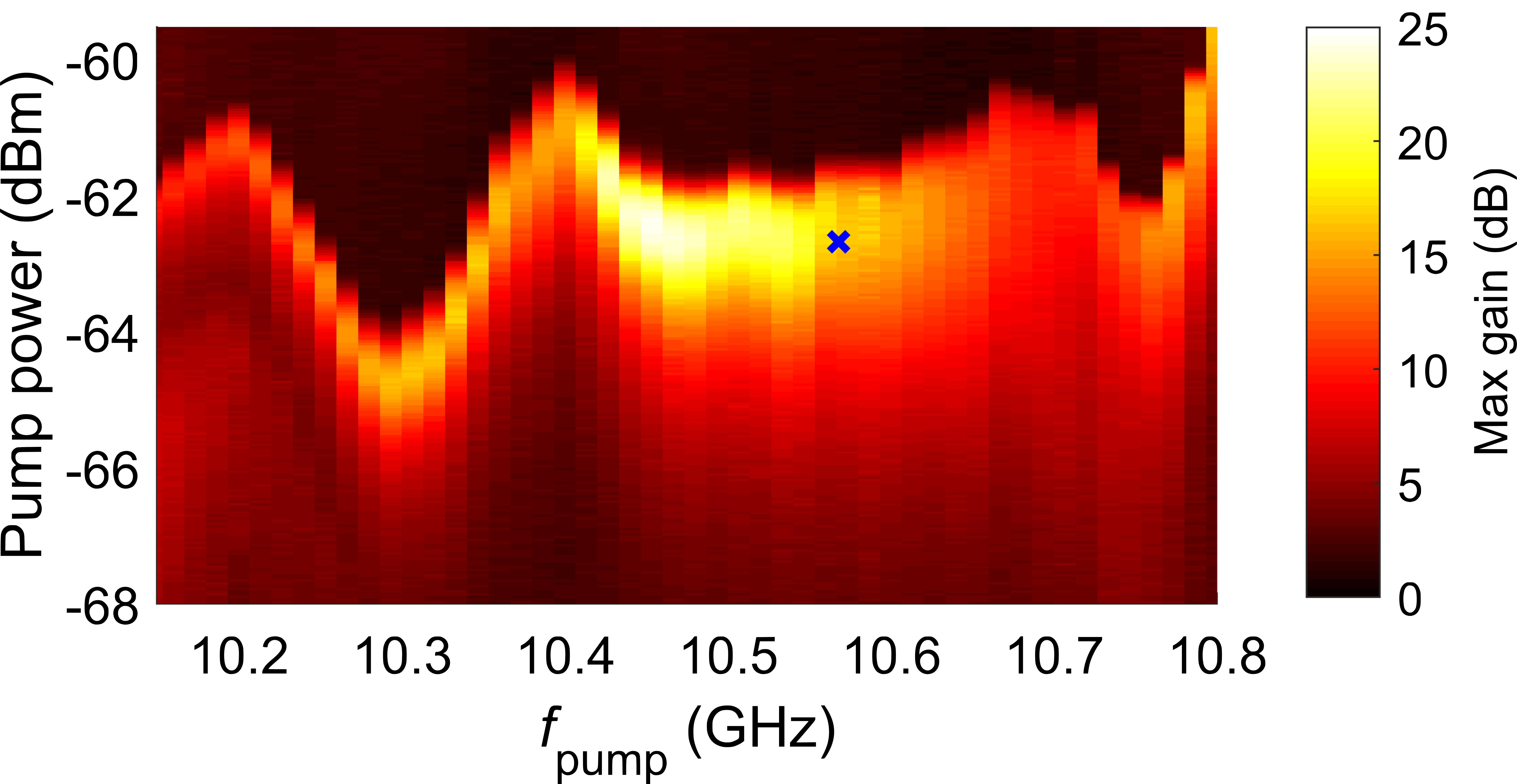}
\caption{Maximum value of the measured gain as a function of pump frequency and power at $\Phi_{\mathrm{DC}} = 0.36\,\Phi_0$. The blue cross denotes the operating point in Fig.~\ref{fig-gain}. \label{fig-surf_gain}}
\end{figure}

We then characterized the JPA gain at various values of $\Phi_{\mathrm{DC}}$. Our JPA exhibited $>$13~dB gain at $0.3\,\Phi_0 < \Phi_{\mathrm{DC}} < 0.4\,\Phi_0$, and the widest tunability was obtained at $\Phi_{\mathrm{DC}} = 0.36\,\Phi_0$. The maximum gain at that point is plotted in Fig.~\ref{fig-surf_gain} as a function of pump frequency and pump power entering the JPA, omitting reflection and finite coupling of the pump tone to the SQUID. The observed variation of the pump power at maximum gain was most likely caused by minor resonances in the flux pumping line, causing the actual RF flux amplitude through the SQUID to fluctuate as a function of the pump frequency at constant applied pump power. It should be noted that the operation frequency can be tuned by several hundred MHz varying the pump frequency alone, as shown in Eq.~(\ref{eq_gain_bw}), while additional tunability from 4.8 to 5.8~GHz can be achieved by varying the DC flux as well.

Noise performance of the JPA was characterized with signal-to-noise ratio improvement method, giving noise relative to system noise temperature, which was calibrated separately using Y-factor method with a heated load.  Thus, we can estimate the noise temperature of the JPA using the following relation\cite{Roy2015a}:
	
	\begin{equation}
		T_{\mathrm{JPA}}(\omega)=T_{\mathrm{tot}}(\omega) \left( \frac{1}{\eta_{SNR}(\omega)}-\frac{1}{G(\omega)} \right),\\[5pt]
	\end{equation}
	
	\noindent where $T_{\mathrm{tot}}(\omega)$ is the total system noise referred to JPA input port including the noise of the HEMT preamplifier and losses in cables and circulators, $\eta_{SNR}(\omega)$ is the SNR improvement and $G(\omega)$ is the gain as defined in Eq.~(\ref{gain}). Our system noise has a mean value of about 6~K.
	
\begin{figure}[!ht]
\centering
\includegraphics[width=0.7\linewidth]{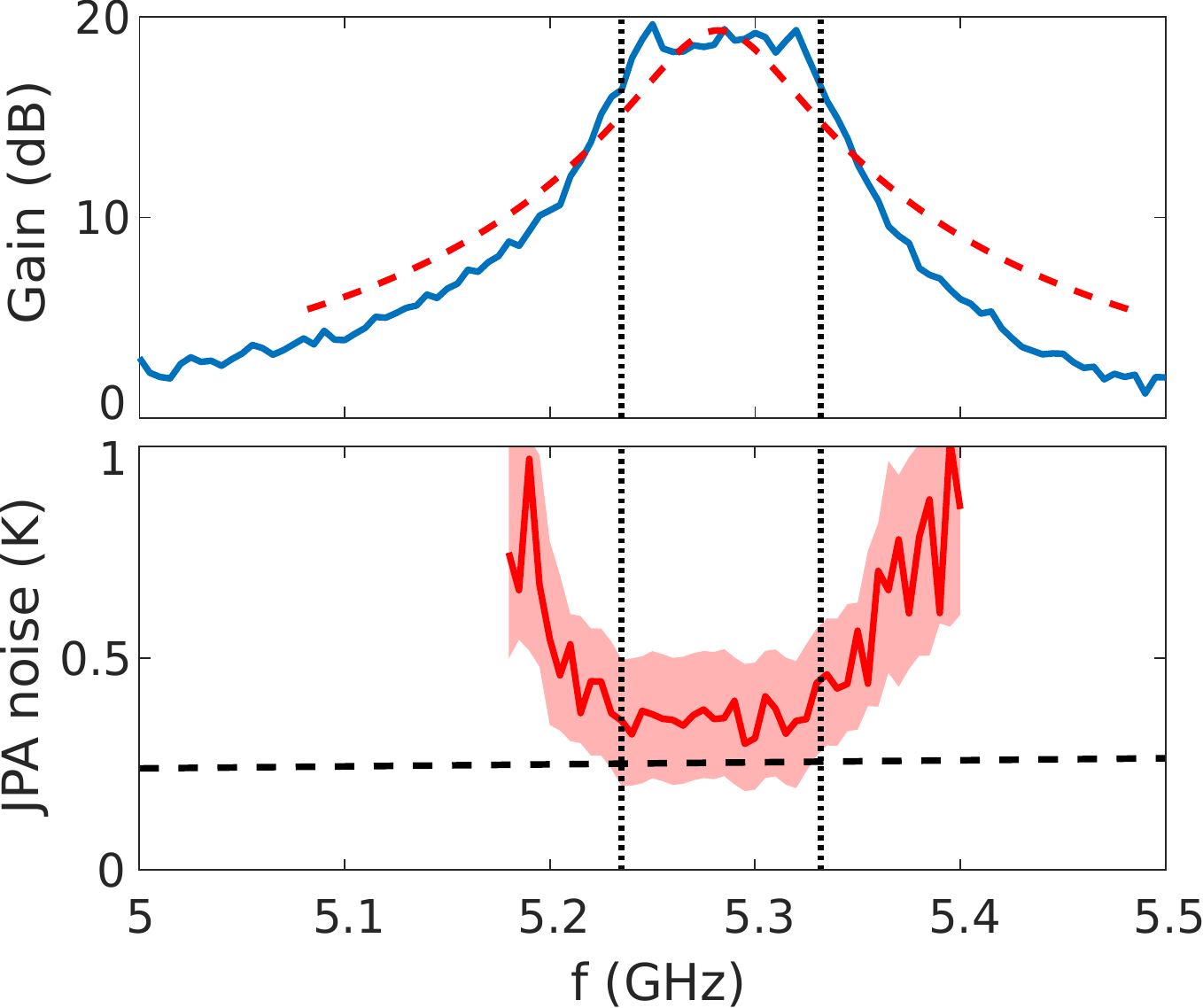}
\caption{Measured gain (upper) and noise (lower) of the JPA at the operation point marked in Fig.~\ref{fig-surf_gain}. The vertical lines denote the -3~dB bandwidth of 95~MHz. % In the gain plot the red dashed line denotes calculated gain.
In the noise plot the shaded area denotes measurement error while the dashed line represents the standard quantum limit ($T_Q = \hbar\omega/k_B$). \label{fig-gain}}
\end{figure}

	The gain and noise characteristics of the JPA in the denoted chosen operating point are shown in Fig.~\ref{fig-gain}. The dashed red line marks the calculated gain outline, showing good agreement with the experiment. In calculations we used $\Phi_{\mathrm{DC}}=0.36\,\Phi_0$ for DC flux and $\Phi_{\mathrm{RF}}=0.07\,\Phi_0$ for RF flux, the strength of pump was $0.45\,\kappa~(= 0.9\,\alpha_{\mathrm{max}})$. %, which corresponds to 90~\% of the value that gives maximum gain.
	We attribute the large error in the noise plot to the inaccuracies in cable losses and system noise temperature measurement. Since our design was optimized for wide bandwidth, the determined input power resulting in 1~dB gain compression was $-125\pm3$~dBm, which, although low, is adequate for various experiments.

%\section{Conclusions}\label{concl}
In summary, we have designed and fabricated a JPA using a straightforward fabrication process with single-step e-beam lithography. The design favoring wide bandwidth over high dynamic range resulted in nearly quantum-limited noise performance with 20~dB gain and 95~MHz bandwidth, which will be useful e.g.\ in shot noise spectroscopy\cite{Nieminen2016} and quantum vacuum measurements\cite{Zhong2013,Lahteenmaki2016}, where signal levels are low. Varying the pump frequency allows rapid tuning of the JPA band center by several hundred MHz. In addition, the bandwidth can be improved further using impedance engineering. Because the JPA is relatively simple to fabricate, the design can be easily modified to meet the requirements of various experimental settings.\\

%\section*{Acknowledgments}
We thank A.\ Lebedev and S.\ Paraoanu for fruitful discussions.
This work was financially supported by the Academy of Finland (projects no. 314448 and 312295), by ERC (grant no. 670743) and by the Ministry of Education and Science of the Russian Federation (grant no. RFMEFI59417X0014). T.E.\ is grateful to V{\"a}is{\"a}l{\"a} foundation of the Finnish Academy of Science and Letters for scholarship. This research work made use of the Aalto
University OtaNano/LTL infrasructure which is part of European Microkelvin Platform. Our project also took advantage of equipment at MIPT Shared Facilities Center.

\end{document}